\documentclass[%
prl,
graphicx,
reprint,
superscriptaddress,
]{revtex4-1}

\usepackage{mathtools} 
\usepackage{amssymb}   
\usepackage{dcolumn}
\usepackage{color}

 \newcommand{\PJ}{$2\,{}^3\!S_1 \rightarrow 2\,{}^3\!P_{\!J}$}
 \newcommand{\Ptwo}{$2\,{}^3\!S_1 \rightarrow 2\,{}^3\!P_2$}

\def\orcid#1{\kern .08em\href{https://orcid.org/#1}{\includegraphics[keepaspectratio,width=0.7em]{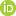}}}

\begin{document}


\title[Collinear Laser Spectroscopy of $^{12}$C$^{4+}$]{Collinear Laser Spectroscopy of \PJ{} transitions in helium-like $^{12}\mathrm{C}^{4+}$} 



\author{P.~Imgram\orcid{0000-0002-3559-7092}}
\email{pimgram@ikp.tu-darmstadt.de}
\affiliation{Institut f\"ur Kernphysik, Departement of Physics, Technische Universit\"at Darmstadt, Schlossgartenstraße 9, 64289 Darmstadt, Germany}

\author{K.~König\orcid{0000-0001-9415-3208}}
\affiliation{Institut f\"ur Kernphysik, Departement of Physics, Technische Universit\"at Darmstadt, Schlossgartenstraße 9, 64289 Darmstadt, Germany}
\affiliation{Helmholtz Research Academy 
Hesse for FAIR, Campus Darmstadt,
Schlossgartenstr.\ 9, 64289 Darmstadt}

\author{B.~Maaß\orcid{0000-0002-6844-5706}}
\altaffiliation{current address: Physics Division, Argonne National Laboratory, 9700 S Cass Ave, IL 60439 Lemont, USA}
\affiliation{Institut f\"ur Kernphysik, Departement of Physics, Technische Universit\"at Darmstadt, Schlossgartenstraße 9, 64289 Darmstadt, Germany}

\author{P.~Müller\orcid{0000-0002-4050-1366}}
\affiliation{Institut f\"ur Kernphysik, Departement of Physics, Technische Universit\"at Darmstadt, Schlossgartenstraße 9, 64289 Darmstadt, Germany}

\author{W.~Nörtershäuser\orcid{0000-0001-7432-3687}}
\affiliation{Institut f\"ur Kernphysik, Departement of Physics, Technische Universit\"at Darmstadt, Schlossgartenstraße 9, 64289 Darmstadt, Germany}
\affiliation{Helmholtz Research Academy 
Hesse for FAIR, Campus Darmstadt,
Schlossgartenstr.\ 9, 64289 Darmstadt}



\date{\today}

\begin{abstract}
Transition frequencies and fine-structure splittings of the \PJ{} transitions in helium-like $^{12}\mathrm{C}^{4+}$ were measured by collinear laser spectroscopy on a 1-ppb level. Accuracy is increased by more than three orders of magnitude with respect to previous measurements, enabling tests of recent non-relativistic QED calculations including terms up to $m\alpha^7$. Deviations between the theoretical and experimental values are within theoretical uncertainties and are ascribed to $m\alpha^8$ and higher-order contributions in the series expansion of the NR-QED calculations. Finally, prospects for an all-optical charge radius determination of light isotopes are evaluated.
\end{abstract}


\maketitle 

\paragraph{Introduction. --}%
In the last decade, advances in experimental precision and sensitivity in laser spectroscopy paired with ab-initio atomic structure calculations have enabled the determination of nuclear charge radii in an all-optical approach purely from optical transition frequencies and related QED calculations. Laser spectroscopy of muonic systems in particular achieved remarkable results in $\mu$H \cite{Pohl2010}, $\mu$D \cite{Pohl2016} and $\mu$He \cite{Krauth2021, Schuhmann2023}. The former suggested a proton radius about 4\% smaller than previously determined from elastic electron scattering and atomic spectroscopy and led to the famous proton-radius puzzle and questioned the lepton universality proposed by the Standard Model of fundamental interactions
(for a recent review see, \textit{e.g.}, \cite{Gao2022}). Although there is convincing evidence for the smaller proton radius as observed in muonic hydrogen from improved measurements of atomic hydrogen \cite{Beyer2017}, a new lamb-shift measurement \cite{Bezginov2019} as well as elastic electron scattering at very small forward-scattering angles using a windowless target \cite{Xiong2019}, some results are still under discussion and atomic hydrogen experiments that support the larger proton radius are also reported \cite{Fleurbaey2018,Brandt2022}. Therefore, an extended comparison between muonic and electronic systems is demanded. However, only the nuclear charge radius of H was studied so far in both systems \cite{Pohl2010, Udem97}.
\newline
In order to expand the all-optical approach towards heavier atomic systems than hydrogen, precise atomic structure calculations for two-electron systems are needed since no laser-addressable narrow electric-dipole transition exists in H-like systems beyond hydrogen. 
Towards this goal, non-relativistic QED (NR-QED) calculations in He-like systems recently made significant progress \cite{Yerokhin2010, Patkos2021, Yerokhin2022Bethe, Yerokhin2022, Yerokhin2023}.
This approach is based on a perturbative expansion series of the level energy in orders $n$ of $m\alpha^n$ with the electron mass $m$ and the fine-structure constant $\alpha$. While the calculated $1s2s\,^3S_1 \rightarrow 1s2p\,^3P_J$ transition energies (in the following text abbreviated as $2\,^3S_1$ and $2\,^3P_J$) are very consistent with experimental data in He, the ionization energies of the individual states as well as the transition energies to the $3\,^{3}\!D$ states differ by up to $10\sigma$ \cite{Patkos2021,Clausen2021,Yerokhin2023}. This could be caused by unknown theoretical contribution shifting the $2\,^{3}\!S$ and $2\,^{3}\!P$ states by about the same amount \cite{Patkos2021}. Since QED contributions scale as $\sim Z^4$, measurements in He-like systems of higher $Z$ have an increased sensitivity and the determination of \PJ{}~transition frequencies and $2\,^3\!P_J$ fine-structure splittings in these ions might provide a hint for the origin of these inconsistencies. So far, only transition frequency measurements in Li$^+$ \cite{Riis94} and Be$^{2+}$ \cite{Scholl93} reached the necessary accuracy to test NR-QED calculations \cite{Yerokhin2022}. While theory and experiment agree for Li$^+$, there is a significant discrepancy in Be$^{2+}$ \cite{Scholl93,Pachucki10,Yerokhin2023}. Helium-like carbon $^{12}$C$^{4+}$ is an excellent candidate for such a test. It is the first stable isotope beyond He that has no nuclear spin and is therefore not plagued by hyperfine-induced fine-structure mixing. Moreover, its nuclear charge radius is accurately known from elastic electron scattering \cite{Cardman80,Sick82,Reuter82,Offermann91} as well as muonic atom spectroscopy \cite{Schaller82,Ruckstuhl84}, yielding consistent results. Thus, spectroscopy of the \PJ{} transitions in He-like carbon provides the ultimate testing ground for higher-order terms in NR-QED calculations as well as for the evaluation of the prospects to extract all-optical nuclear charge radii from He-like ions beyond He.     
\newline
\begin{figure*}[htb]%
\centering
\includegraphics[width=\linewidth]{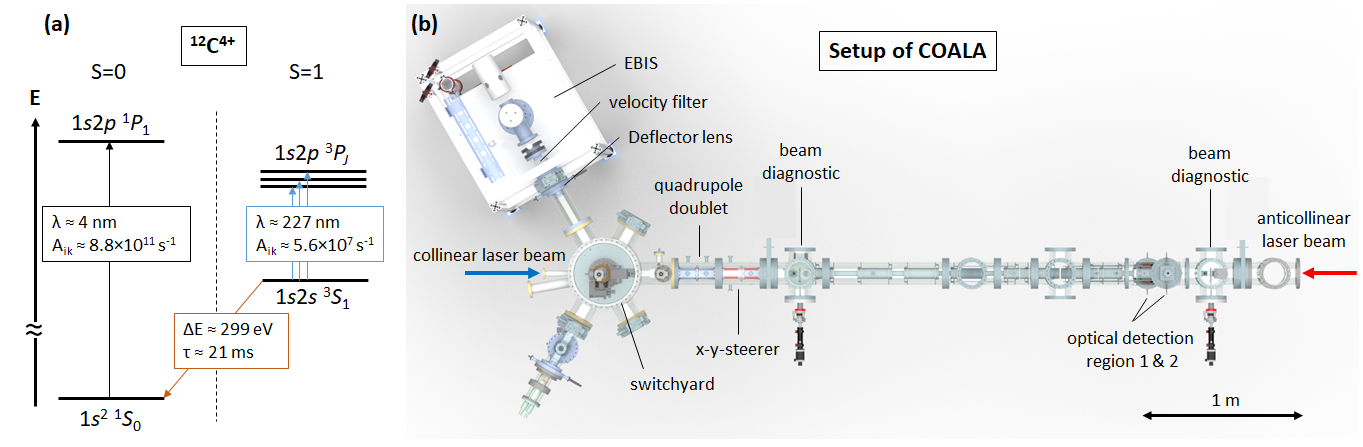}
\caption{(a) Level scheme of $^{12}$C$^{4+}$. (b) Experimental setup of the COALA beamline. Highly charged $^{12}$C$^{4+}$ ions are produced in an electron beam ion source (EBIS) on a 10.5\,kV starting potential. After acceleration of the ions into the beamline, they are superposed with a collinear and an anticollinear laser beam to perform quasi-simultaneous collinear and anticollinear laser spectroscopy in the optical detection region where the fluorescence light is collected and detected by two elliptical mirrors and photomultiplier tubes.}\label{fig:setup}
\end{figure*}
In this work, we present \PJ{}~transition frequency measurements in $^{12}$C$^{4+}$ with 1.3 part-per-billion (ppb) accuracy which represents an improvement by three orders of magnitude compared to previous experiments. Our results surpass the theoretical accuracy by roughly two orders of magnitude and therefore become sensitive to higher-order QED terms in NR-QED beyond He.\\

\paragraph{Method. --}%
The level structure of He-like ions can be subdivided in singlet ($S=0$) and triplet ($S=1$) states where the electron spins are aligned anti-parallel or parallel, respectively. As illustrated in Fig.\,\ref{fig:setup}(a), in $^{12}$C$^{4+}$ no laser-accessible transition exists from the $1s^2\,^1\!S_0$ ground state. However, the metastable $2\,^3\!S_1$ triplet state can serve as the lower state for laser excitation into the $2\,^3\!P_J$ states at a laser wavelength of $\lambda \approx 227$\,nm if it is sufficiently populated in the ion source. Here, we use an electron beam ion source (EBIS) which is well known to produce highly charged ions and to populate metastable states. But since spectroscopy inside an EBIS is limited by the very high temperature and the corresponding Doppler  broadening even if forced evaporative cooling is applied \cite{Maeckel2011}, it is necessary to transport the ions into an environment that is more appropriate for high-resolution spectroscopy. This can boost accuracy by several orders of magnitude as it has been demonstrated for Ar$^{13+}$ ions in a Penning \cite{Egl2019} or a Paul trap \cite{Micke2020, King2022}. However, the short lifetime of the $2\,^3\!S_1$ state in He-like C$^{4+}$ ions ($\tau \approx 21$\,ms) requires a fast technique like collinear laser spectroscopy, which had originally been developed to perform spectroscopy of short-lived isotopes \cite{Schinzler78}.
This was realized by coupling an EBIS (DREEBIT, EBIS-A) to the Collinear Apparatus for Laser Spectroscopy and Applied Physics (COALA) \cite{König20COALAReview} situated at the Institute for Nuclear Physics of TU Darmstadt as depicted in Fig.\,\ref{fig:setup}(b). At COALA, we achieved accuracies of the order of 100\,kHz in previous studies on singly charged ions \cite{Imgram19, Müller20, König2020}, comparable to ion-trap measurements of allowed dipole transitions \cite{Herrmann2009,Batteiger2009,Gebert2015,Shi2016}. The He-like ions are produced through electron impact ionization and a significant fraction ($\approx 40\%$) of the ions is transferred into the metastable state almost exclusively through charge exchange from C$^{5+}$ to C$^{4+}$. The initial energy spread strongly depends on the production parameters as detailed in \cite{Imgram23_PRA} and was $1.1\,\mathrm{eV}$ in the applied continous-beam mode. Then, the ions are accelerated from their starting potential $U_\mathrm{start} \approx 10.5$\,kV towards ground potential into the beamline. The choice of $U_\mathrm{start}$ was a compromise of having a high starting potential for maximum velocity compression while limiting recurrences of discharges inside the source. Through subsequent electrostatic ion optics and deflectors, the ion beam is superposed with two laser beams, of which one is copropagating with  the ion beam and the other one counterpropagating. By alternately blocking one of the two laser beams, we perform frequency‑comb referenced quasi-simultaneous collinear and anticollinear laser spectroscopy as it was previously applied for short-lived Be isotopes   \cite{Nörtershäuser2009, Krieger2016}. The resonance fluorescence signal is recorded with photomultiplier tubes (PMT) attached to the optical detection region that can be floated to a variable potential to realize Doppler tuning. Hereby, the Doppler-shifted laboratory-frame transition frequencies
$
    \nu_\mathrm{c/a} = \nu_0 \gamma (1 \pm \beta)
$ 
in collinear ($\nu_\mathrm{c}$) and anticollinear ($\nu_\mathrm{a}$) geometry are measured in fast iterations. The rest-frame transition frequency $\nu_0$ is extracted as the geometrical average 
\begin{equation}
    \nu_\mathrm{c}\nu_\mathrm{a} = \nu_0^2 \gamma^2 (1 + \beta)(1 - \beta) 
    = \nu_0^2
    \label{eq:colAcol}
\end{equation}
without the need of the knowledge of the precise ion velocity $\beta = \upsilon/c$ and the corresponding Lorentz factor $\gamma = 1/\sqrt{1 - \beta^2}$. This reduces the systematic uncertainties, which are then dominated by the uncertainty stemming from the alignment of the laser beams as discussed in detail in \cite{Imgram23_PRA}.
\newline
The laser system \cite{König20COALAReview} consists of two continuous-wave titanium:sapphire (Ti:Sa) lasers (Matisse 2), each pumped by a frequency-doubled Nd:YAG laser (Millennia eV20). The emitted 908\,nm light is frequency quadrupled and transported in free space to the beamline. Lens telescopes are used to achieve a collimated laser beam in both directions with a beam diameter of about 1\,mm inside the optical detection region. The laser power of both lasers was held constant at 0.5\,mW during the measurements. A GPS-disciplined quartz oscillator provided the 10\,MHz reference for the Menlo-Systems FC1500-250-WG frequency comb used to measure and stabilize the fundamental laser frequency. The resulting laser linewidth of the fundamental laser light was approximately 200\,kHz. The laser light had linear polarization to suppress potential Zeeman shifts induced by external magnetic fields \cite{Imgram23_PRA}.

\begin{figure*}[htb]%
\centering
\includegraphics[trim=0mm 0mm 0mm 0mm,clip,width=1.0\linewidth]{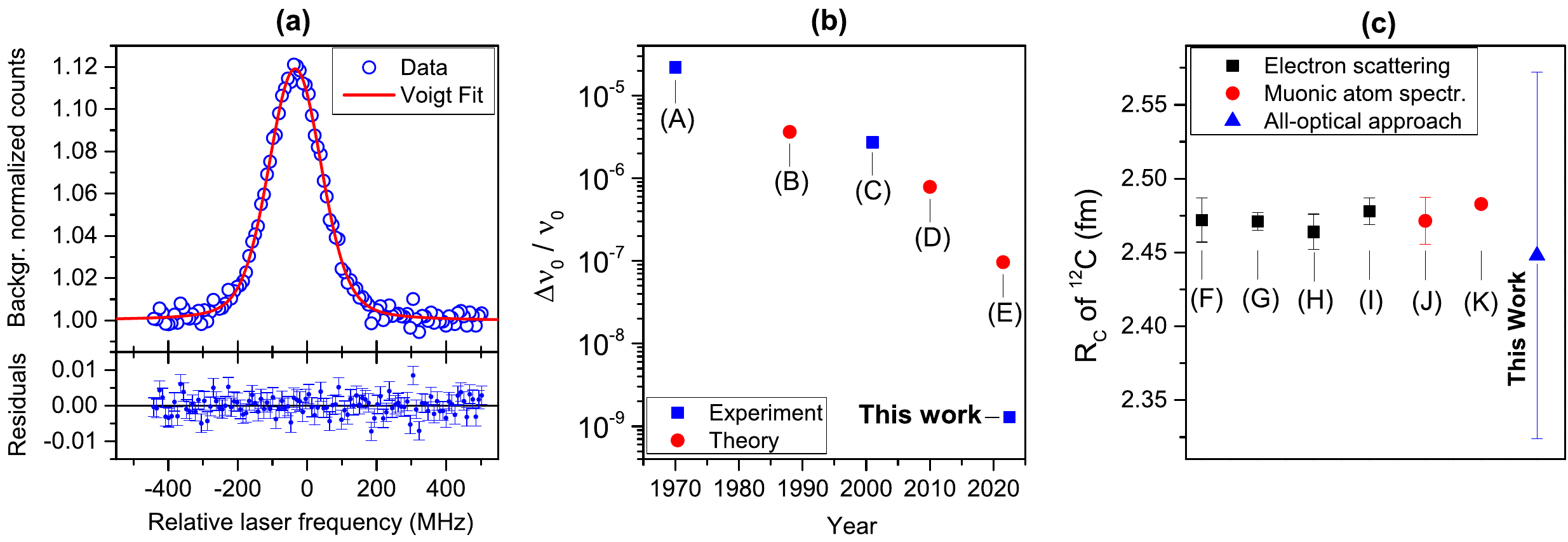}
\caption{(a) Background-normalized anticollinear resonance spectrum of the \Ptwo ~transition in $^{12}$C$^{4+}$ with a Voigt profile fit (red line) and the respective residuals. (b) Relative precision $\Delta \nu_0 / \nu_0$ of literature values (A) - (E) from Tab.\,\ref{tab:freq_comp} for the \Ptwo ~transition as a function of time including our result. (c) Literature values (F) - (K) of the nuclear charge radius of $^{12}$C from Tab.\,\ref{tab:charge_radius} in comparison to the all-optical nuclear charge radius extracted here.}\label{fig:results}
\end{figure*}

\paragraph{Transition frequencies. --}%
A typical resonance spectrum depicted in Fig.\,\ref{fig:results}(a) is obtained by plotting the PMT counts as a function of the laser frequency. To visualize the signal-to-background ratio, the photon counts are normalized to the background rate. In order to extract the center frequency $\nu_\mathrm{c/a}$ from the spectrum, a Voigt profile, which is the convolution of a Gaussian and Lorentzian profile, was fitted to the data. The resulting full width at half maximum (FWHM) of roughly 170\,MHz is dominated by the Gaussian contribution which originates from the energy spread of the ions \cite{Imgram23_PRA}. The choice of the fitting lineshape does not shift the result significantly as long as the same profile is used for both directions. Each pair consisting of a collinear and an anticollinear measurement yields a rest-frame frequency through Eq.\,(\ref{eq:colAcol}). The final frequencies $\nu_0 (\,^3\!P_J)$ of the \PJ{} transitions were determined by taking the average over all pairs weighted by their statistical fitting uncertainties. Results are listed in Tab.\,\ref{tab:freq_comp} and compared with previous experimental values and theoretical predictions.\\ 
\begin{table*}[htb]
    \centering
    \caption{Comparison between literature values and this work of the rest-frame frequencies $\nu_0(\,^3\!P_J)$ in the \PJ{} transitions of $^{12}$C$^{4+}$. The values in brackets denote the full $1\sigma$ uncertainty. All values are in GHz.}
    \begin{ruledtabular}
        \begin{tabular}{c  d  d  d  l}
	    \multicolumn{1}{c}{Label Fig.\,\ref{fig:results}} &\multicolumn{1}{c}{$\nu_0 (\,^3\!P_2)$} & \multicolumn{1}{c}{$\nu_0 (\,^3\!P_1)$} & \multicolumn{1}{c}{$\nu_0 (\,^3\!P_0)$} & \multicolumn{1}{c}{Ref.}\\
	    \hline
        (A) & 1\,319\,747.\,(29)    & 1\,315\,669.\,(29)  & 1\,316\,056.\,(29) & [\onlinecite{Edlen1970}] (Exp.)\\
        (B) & 1\,319\,744.6\,(4.8) & 1\,315\,076.2\,(4.8) & 1\,316\,057.7\,(4.8) & [\onlinecite{Drake88}] (Theory) \\
        (C) & 1\,319\,753.0\,(3.6) & 1\,315\,679.7\,(4.2) & 1\,316\,056.2\,(4.5) & [\onlinecite{Ozawa01}] (Exp.)\\
        (D) & 1\,319\,748.6\,(1.0) & 1\,315\,677.5\,(1.0) & 1\,316\,052.5\,(1.0) & [\onlinecite{Yerokhin2010}] (Theory)\\
        (E) & 1\,319\,748.55\,(13) & 1\,315\,677.60\,(75) & 1\,316\,052.03\,(27) & [\onlinecite{Yerokhin2022}] (Theory)\\
        \hline
        ~ & 1\,319\,748.571\,4\,(17) & 1\,315\,677.192\,8\,(17) & 1\,316\,052.219\,3\,(19) & This work\\
        \end{tabular}
        \end{ruledtabular}
    \label{tab:freq_comp}
\end{table*}
Statistical uncertainties are estimated as the standard error of the mean for 108 ($^3\!P_2$), 68 ($^3\!P_1$), and 28 ($^3\!P_0$) measurements that were taken over several weeks. Our total systematic uncertainty of the rest-frame transition frequency is $\sim$1.7\,MHz, dominated by the remaining ion beam divergence in combination with the superposition of the two laser beams that might lead to probing slightly different velocity classes in both directions (for details see \cite{Imgram23_PRA}). 
The systematic uncertainty is added in quadrature to the statistical uncertainty and a combined 1$\sigma$ uncertainty of less than 2\,MHz is obtained for all transitions. This represents an improvement of more than three orders of magnitude compared to previous experimental results in $^{12}$C$^{4+}$ \cite{Ozawa01} as illustrated in Fig.\,\ref{fig:results}(b). Here, the relative accuracy $\Delta \nu_0 / \nu_0$ in experiment (blue) and theory (red) is plotted on a logarithmic axis as a function of time. A steady improvement in accuracy is visible with a substantial gain through our work. Our experimental results agree very well with recent NR-QED calculations \cite{Yerokhin2022} whose uncertainties are, however, two orders of magnitude larger than our experimental uncertainties.

\paragraph{Fine-structure splitting. --}%
Another important test of theory is the fine-structure splitting of the $2\,^3\!P_J$ states. While the theoretical accuracy of the \PJ ~transition frequencies in $^{12}$C$^{4+}$ is currently limited to $\sim$130\,MHz \cite{Yerokhin2022}, the fine-structure splittings in $^{12}$C$^{4+}$ were calculated on the $\sim$10\,MHz level \cite{Pachucki10} including $m\alpha^7$ terms, but could never be tested hitherto because of the lack of experimental data. 
The fine structure splittings can be directly obtained from the differences of our measured transition frequencies and
the values are listed in Tab.\,\ref{tab:fine_struc_split} together with the theoretical predictions by Pachucki and Yerokhin \cite{Pachucki10}. All values agree within the stated theoretical uncertainties. The difference between the experimental and the theoretical value represents the sum of all higher-order contributions and are, thus, an approximation of the dominant $m\alpha^8$ terms. Their contribution was estimated by scaling the calculated $m\alpha^6$ contribution by
the factor $(Z\alpha)^2$ and used to represent the theoretical uncertainty \cite{Yerokhin2022}. In atomic helium the corresponding experimental values \cite{Smiciklas2010,Zheng2017} were in excellent agreement with the theoretical result and, thus, the $m\alpha^8$ contribution to the splitting between the unmixed $J=0$ and $J=2$ levels were expected to be small. Confirming the overestimation of the $m\alpha^8$ contributions in the more sensitive case of $^{12}$C$^{4+}$ would have allowed to revise the uncertainty estimation and, thus, to improve the uncertainty of the fine-structure constant $\alpha$ \cite{Pachucki10}. However, our experimental result indicates a rather large contribution of higher orders. Instead, our result can now be used as a benchmark for calculations that approximate the $m\alpha^8$ contributions to identify the dominant terms, which finally can also lead to an improved He fine-structure constant.

Finally, we note the larger difference between experiment and theory for the $^3\!P_{0}-{^3\!P_{1}}$ interval compared with the $^3\!P_{0}-{^3\!P_{2}}$ interval. This is most probably due to the singlet-triplet configuration mixing of the $2\,^3\!P_1$ and the $2\,^1\!P_1$ states having the same total angular momentum and parity and are close in energy \cite{Yerokhin2022}. This mixing is enhanced with increasing $Z$ as compared to other $m\alpha^8$ effects which is also reflected in the larger uncertainty from theory.

\begin{table}[!ht]
    \centering
    \caption[Fine-structure splitting of the $2\,^3\!P_J$ levels in $^{12}$C$^{4+}$]{Fine-structure splitting of the $2\,^3\!P_J$ levels in $^{12}$C$^{4+}$ acquired from this work and theory \cite{Pachucki10}. All values are in MHz.}
    \begin{ruledtabular}
        \begin{tabular}{c d d}
	    $(J,J')$ & \multicolumn{1}{c}{$(0,1)$} & \multicolumn{1}{c}{$(0,2)$}\\\hline
        This work & -375\,026.5(2.5) & 3\,696\,352.1(2.5)\\
        Theory \cite{Pachucki10} & -374\,996.3\,(48.0) & 3\,696\,343.5\,(10.0)\\\hline
        Difference & -30.2\,(48.0) & 8.6\,(10.0)\\
        \end{tabular}
    \end{ruledtabular}
    \label{tab:fine_struc_split}
\end{table}

\paragraph{Charge Radii. --}%
By confirming the NR-QED calculations in two-electron systems, these can be combined with the experimental data to determine absolute nuclear charge radii in an all-optical approach. The measured transition frequency $\nu_0$ of the \Ptwo\ transition can be written as
\begin{equation}
    \nu_{0} = \nu_\mathrm{point} + F \cdot {R_\mathrm{c}^2}
    \label{eq:all-optical}
\end{equation}
where $\nu_\mathrm{point}$ is the calculated transition frequency assuming a point-like but finite-mass nucleus and $F$ the calculated field-shift factor of the transition. Both values are calculated by Yerokhin \textit{et al.} but not explicitly tabulated in \cite{Yerokhin2022}. Instead, we obtained $\nu_\mathrm{point} =1\,319\,749.83(13)$\,GHz as the difference of the ionization energies of the $2\,^3\!S_1$ and the $2\,^3\!P_2$ levels after subtracting the nuclear size contribution given also in the tables. Dividing the latter by $R_\mathrm{c}^2$ using $R_\mathrm{c}(^{12}\mathrm{C}) = 2.4702(22)$\,fm as applied in \cite{Yerokhin2022} provides $F= 0.2115$\,GHz/fm$^2$. Using these values gives $R_\mathrm{c}(^{12}\mathrm{C}) = 2.45\,(12)\,\mathrm{fm}$ in excellent agreement with previous results as listed in Tab.\,\ref{tab:charge_radius}. 
In case that $\Delta\nu_\mathrm{point}$ can be reduced to the size of $\Delta\nu_{0,\mathrm{exp}}$, the charge radius could be directly obtained with an unprecedented accuracy of $\pm 0.0016$\,fm, which is even more accurate than the determination from muonic atom spectroscopy \cite{Ruckstuhl84}. 
While this is a daunting task for theory, we note that for the stable isotopes of boron, already an improvement of the theoretical uncertainty by a factor of 2--3 would allow for an improved charge radius determination since they are only poorly known from elastic electron scattering and muonic atom spectroscopy \cite{Maaß2019}. A more reliable charge radius of stable boron isotopes is of particular  importance for the ongoing campaign to determine the charge radius of the proton-halo candidate $^8$B \cite{Maass2017}.
 
\begin{table}[!h]
    \centering
    \caption{Nuclear charge radius of $^{12}$C. All values are in fm.}
    \begin{minipage}{0.31\textwidth}
    \begin{ruledtabular}
        \begin{tabular}{clc}
	    {~} & {R$_\mathrm{c}$($^{12}$C)} & {Ref.}\\\hline
        (F) & 2.472(15) & [\onlinecite{Cardman80}]\\
        (G) & 2.471(6) & [\onlinecite{Sick82}] \\
        (H) & 2.464(12) & [\onlinecite{Reuter82}]\\
        (I) & 2.478(9) & [\onlinecite{Offermann91}]\\
        (J) & 2.472(16) & [\onlinecite{Schaller82}]\\
        (K) & 2.4829(19)& [\onlinecite{Ruckstuhl84}] \\
        \hline
        ~ & 2.45(12) & This work\\
        \end{tabular}
        \end{ruledtabular}
    \label{tab:charge_radius}
    \end{minipage}
\end{table}

\paragraph{Summary. --}%
We measured the transition frequencies of all fine-structure components in the \PJ{} transitions in He-like $^{12}$C$^{4+}$ ions with 1.3\,ppb accuracy and confirmed NR-QED calculations of these transitions within their uncertainties. We combined the production of He-like ions in the metastable $2\,^3\!S_1$ state in an EBIS with collinear laser spectroscopy and demonstrated frequency determinations at the 1-MHz level. The experimentally determined fine-structure splittings of the $^3\!P_J$ states pave the way to further decrease
the theoretical uncertainty of the fine-structure constant, required for constraining the fine-structure constant in He. Based on the NR-QED calculations, the charge radius of $^{12}$C was extracted directly from the optical transition frequency with an uncertainty of $\sim$5\%, only limited by the theoretical accuracy. The technique will be applied next to He-like $^{13}$C ions. Measuring the hyperfine splitting and isotope shift with respect to $^{12}$C, a very accurate charge radius of $^{13}$C can be obtained based on conventional mass-shift calculations as they have been used for determinations of charge radii of He, Li, Be \cite{Lu2013} and B isotopes \cite{Maaß2019}.\\
Finally, measurements on He-like and Li-like boron ions and neutral boron \cite{Maaß2019} will provide a very precise differential charge radius between the B isotopes allowing to test the consistency of NR-QED calculations in two-, three-, and five-electron systems.

\begin{acknowledgments}
\paragraph{Acknowledgment. --}
We thank J.~Krämer for his contributions in the early stage of the project and K.~Pachucki \& V.~Yerokhin for many fruitful discussions. We acknowledge funding by the Deutsche Forschungsgemeinschaft (DFG) Project No. 279384907 SFB 1245, as well as under Grant INST No.\ 163/392-1 FUGG, and from the German Federal Ministry for Education and Research (BMBF) under Contract No. 05P21RDFN1. P.I. and P.M. acknowledge support from HGS-HIRE.
\end{acknowledgments}

\end{document}